\documentclass[11pt,preprint]{aastex}
\usepackage{epsfig}
\usepackage{natbib}
\usepackage{graphicx}
\usepackage{slashbox}
\usepackage{multirow}
\usepackage{lscape}
\usepackage{mathrsfs,amssymb}
\usepackage{amsmath}
\newcommand       \Angstrom     {\,{\rm \AA}}

\newcommand       \cm           {\,{\rm cm}}

\newcommand       \K            {\,{\rm K}}

\newcommand       \gtsim        {\gtrsim}

\newcommand       \mum          {\,{\rm \mu m}}

\newcommand       \simali       {\sim\,}

\newcommand       \rmH          {{\rm H}}
\shorttitle{DIBs vs. the 2175$\Angstrom$ Extinction Bump}
\title{
\vspace*{-2.0em}
{\normalsize\rm {\it The Astrophysical Journal Letters}, in press}\\
\vspace*{1.0em}
The Carriers of the Interstellar Unidentified Infrared Emission Features: 
Aromatic or Aliphatic?
}

\author{Aigen Li\altaffilmark{1} and B.T. Draine\altaffilmark{2}} 
\altaffiltext{1} {Department of Physics and Astronomy,
                  University of Missouri,
                  Columbia, MO 65211, USA;
                  {\sf lia@missouri.edu}}
\altaffiltext{1} {Princeton University Observatory, 
                  Princeton University, 
                  Princeton, NJ 08544, USA; 
                  {\sf draine@astro.princeton.edu}}

\begin{document}

\begin{abstract}
The unidentified infrared emission (UIE) features 
at 3.3, 6.2, 7.7, 8.6, and 11.3$\mum$,
commonly attributed to polycyclic aromatic hydrocarbon 
(PAH) molecules,  
have been recently ascribed to
coal- or kerogen-like organic nanoparticles 
with a mixed aromatic-aliphatic structure.
However, we show in this {\it Letter} that this hypothesis 
is inconsistent with observations. 
We estimate the aliphatic fraction of the UIE carriers 
based on the observed intensities of 
the 3.4$\mum$ and 6.85$\mum$ emission features
by attributing them exclusively to 
aliphatic C--H stretch and aliphatic
C--H deformation vibrational modes, respectively. 
We derive the fraction of carbon atoms
in aliphatic form to be $<$\,15\%.
We 
conclude that the UIE emitters are predominantly
aromatic,
with 
%
aliphatic material
at most a minor part
of the UIE carriers. The PAH model is consistent with astronomical
observations and PAHs 
dominate
the 
strong
UIE bands.
\end{abstract}

\section{Introduction\label{sec:intro}}
The ``unidentified infrared emission'' (UIE) bands, 
a distinct set of spectral features at wavelengths 
of 3.3, 6.2, 7.7, 8.6, 11.3 and 12.7$\mum$, 
dominate the mid-infrared spectra of many bright 
astronomical objects. They are ubiquitously seen in 
the interstellar medium (ISM) of our own galaxy and 
star-forming galaxies, both near and far, 
and account for over 10\% of their total infrared (IR) 
luminosity (see Joblin \& Tielens 2011). 
Although the exact nature of the carriers 
remains unknown, 
the UIE bands
are commonly attributed 
to polycyclic aromatic hydrocarbon (PAH) molecules 
(L\'eger \& Puget 1984, Allamandola et al.\ 1985). 
The identification of the UIE bands is important as 
they are a useful probe of the cosmic star-formation history, 
and their carriers are an essential player in galactic evolution.

Very recently, Kwok \& Zhang (2011; hereafter KZ11) 
argue
that the UIE bands 
arise from coal- or kerogen-like organic nanoparticles,
consisting 
of
chain-like aliphatic hydrocarbon material linking
small units of aromatic rings.
This hypothesis 
has potentially
important implications for our understanding of 
stellar evolution, interstellar chemistry, and the formation of 
our solar system. If confirmed, 
it
would establish an important 
link among stars at their late evolutionary stages, the ISM, 
and the solar system, as the kerogen-like organic matter seen 
in meteorites (Derenne \& Robert 2010; Cody et al.\ 2011)
has similar chemical structures as those suggested 
by KZ11 for the UIE bands seen in the ISM 
and in circumstellar environments around evolved stars 
(i.e., planetary nebulae and proto-planetary nebulae).

However, the KZ11 hypothesis of substantially 
aliphatic organic matter as the UIE carriers 
does not appear to be consistent with the observed strengths of the UIE bands.
As will be elaborated below in \S\ref{sec:ch3.4um} 
and \S\ref{sec:ch6.85um}, astronomical observations show that if 
aliphatic hydrocarbon units are present in the UIE carriers, 
they must be a minor constituent. 
Further, their arguments against the PAH model 
do not seem to pose a problem (see \S\ref{sec:discussion}).  

\section{Constraints on the Aliphatic Fraction 
%
         from the 3.4$\mum$ Feature\label{sec:ch3.4um}}
%
KZ11
argue that the material responsible for 
the UIE features has a substantial aliphatic component, 
based on the mid-IR spectra of NGC\,7027 (a planetary nebula), 
IRAS\,22272+5435 (a protoplanetary nebula), and the Orion bar 
(a photodissociation region in the Orion nebula). 
They decompose the 3--20$\mum$ spectra of these objects 
into three components: the UIE bands, 
broad plateaus (several $\mum$ in width) 
peaking at 8 and 12$\mum$, and a thermal continuum. 
They attribute the broad plateau features 
(which account for $\simali$1/3 of the 3--20$\mum$ power 
of these objects) to aliphatic branches of the UIE carriers, 
similar to the coal model for the UIE bands (Guillois et al.\ 1996).
Recognizing the challenge of the coal model in being 
%
heated to emit the UIE bands 
(Puget et al.\ 1995), 
KZ11
hypothesize 
that the coal-like UIE carriers are nanometer in size 
or they are heated by the chemical energy released from 
the $\rmH$\,$+$\,$\rmH$\,$\rightarrow$\,$\rmH_2$ reaction
(Duley \& Williams 2011).

Aliphatic hydrocarbon has a 
band at 3.4$\mum$ due to the C--H stretching
mode
(Pendleton \& Allamandola 2002). 
In some HII regions, reflection nebulae and planetary nebulae
(as well as extragalactic regions, 
e.g., see Yamagishi et al.\ 2012, Kondo et al.\ 2012), 
the UIE 
%
near 3$\mum$ exhibits a rich spectrum: 
the dominant 3.3$\mum$ feature is usually accompanied 
by a weaker feature at 3.4$\mum$ 
along with an underlying plateau 
extending out to $\simali$3.6$\mum$.
In some objects, a series of weaker features 
at 3.46, 3.51, and 3.56$\mum$ are also seen superimposed 
on the plateau, showing a tendency to decrease in strength 
with increasing wavelength (see Figure~1
and Geballe et al.\ 1985, Jourdain de Muizon et al.\ 1986,
Joblin et al.\ 1996).
While assignment of the 3.3$\mum$ emission feature to 
the aromatic C--H stretch is widely accepted, 
the precise identification of the 3.4$\mum$ feature 
(and the accompanying weak features at 3.46, 3.51, and 3.56$\mum$ 
and the broad plateau) remains somewhat controversial. 
By assigning the 3.4$\mum$ emission exclusively 
to aliphatic C--H, one can place an upper limit 
on the aliphatic fraction of the emitters of the UIE features.

Let $I_{3.4}$ and $I_{3.3}$ respectively be the observed intensities 
of the 3.4$\mum$ and 3.3$\mum$ emission features. 
In interstellar and circumstellar environments, 
$I_{3.4}/I_{3.3}$ typically ranges from $\simali$0.06 
to $\simali$0.20, depending on the local conditions
(Schutte et al.\ 1993).
Let $A_{3.4}$ and $A_{3.3}$ respectively be the band strengths 
of the aliphatic and aromatic C--H bonds. 
We take $A_{3.4} = 2.7\times10^{-18}\cm$ per aliphatic C--H bond, 
averaged over ethane, hexane, ethyl-benzene, 
and methyl-cyclo-hexane (d'Hendecourt \& Allamandola 1986, 
Mu\~noz-Caro et al.\ 2001).\footnote{%
  Typical type II Kerogens have
  $A_{3.4} \approx 2.8\times10^{-18}\cm$ per C atom 
  while the 3.3$\mum$ aromatic feature is barely visible
  (see Figure~2 in Papoular 2001). This clearly shows that 
  kerogen -- at least this type -- is not a good explanation 
  for the 3.3$\mum$ and 3.4$\mum$
  emission
  features. 
  In coals, the 3.3$\mum$ aromatic feature is usually
  weaker than the 3.4$\mum$ aliphatic feature
  except for those with high ranks (i.e., more evolved,
  more ordered, with lower H/C and O/C ratios).
  As coal evolves, the progressive release of heteroatoms 
  (decreasing
  H/C and O/C) leads to 
  formation 
  of planar clusters of benzene-type rings followed by 
  stacking of these aromatic sheets to form disordered stacks of 
  graphitic planes (see Papoular 2001).
  As a result of the progressive aromatization,
  $A_{3.4}/A_{3.3}$ decreases as coal evolves. 
  The aliphatic C--H deformation band at 6.85$\mum$
  band disappears in highly evolved coal 
  (e.g., anthracite, see Papoular 2001). 
  %
  }
We take $A_{3.3} = 4.0\times10^{-18}\cm$ per aromatic C--H bond 
for small neutral PAHs (Draine \& Li 2007).

\begin{figure}
\centerline
{
\includegraphics[width=8cm,angle=0]{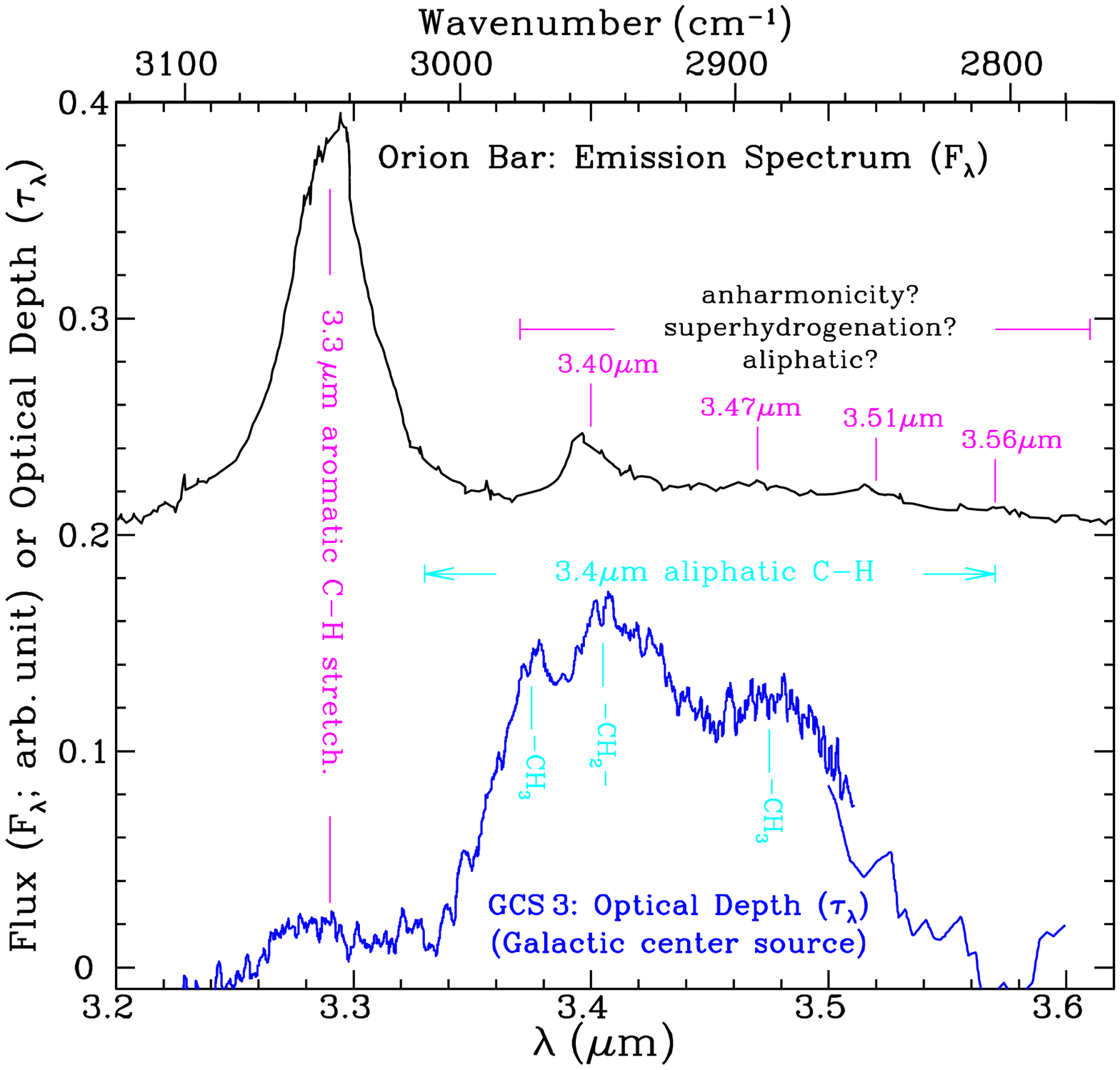}
}
\caption{\footnotesize
         \label{fig:1}
         Comparison of the 3.15--3.65$\mum$ {\it emission} spectrum 
         of the Orion Bar (position 4; black; Sloan et al.\ 1997) 
         with the optical  depth ({\it absorption}) spectra of GCS\,3 
         (a Galactic center source; blue; Chiar et al.\ 2000). 
         The weakness of the 3.4$\mum$ 
feature in the {\it emission}
         spectrum indicates that 
         the {\it aliphatic} component must be minor, 
         even assuming that the 3.4$\mum$ emission is exclusively 
         due to aliphatic C--H (i.e., neglecting anharmonicity 
         and superhydrogenation). In contrast, the 
         {\it absorption} spectrum
         of the diffuse ISM
toward GCS\,3 is
         dominated by aliphatic hydrocarbon.
         }
\end{figure}


Let $N_{\rm H,aliph}$ and $N_{\rm H,arom}$ respectively 
be the numbers of aliphatic and aromatic C--H bonds 
in the emitters of the 3.3$\mum$ UIE feature. 
We obtain 
$N_{\rm H,aliph}/N_{\rm H,arom}\approx\left(I_{3.4}/I_{3.3}\right)
\times\left(A_{3.3}/A_{3.4}\right)\approx0.30$, 
taking $I_{3.4}/I_{3.3}$\,=\,0.2
[KZ11
estimate $I_{3.4}/I_{3.3}\approx0.22$
for NGC\,7027, and $I_{3.4}/I_{3.3}\approx0.19$ for the Orion bar].
We assume that one aliphatic C atom corresponds to 
2.5 aliphatic C--H bonds (intermediate between methylene -CH$_2$ 
and methyl -CH$_3$) and one aromatic C atom corresponds to 
0.75 aromatic C--H bond (intermediate between benzene C$_6$H$_6$ 
and coronene C$_{24}$H$_{12}$). 
Therefore, in the UIE carriers the ratio of the number of C atoms 
in aliphatic units to that in aromatic rings is 
$N_{\rm C,aliph}/N_{\rm C,arom}\approx 0.30\times\left(0.75/2.5\right) 
= 0.09$, showing that the aliphatic component is only a minor part of 
the UIE emitters. 
%
KZ11
take $I_{3.4}/I_{3.3}\approx1.88$ 
for the protoplanetary nebula IRAS\,22272+5435 
(but much smaller $I_{3.4}/I_{3.3}$ ratios have also been 
reported for this source; see Goto et al.\ 2003). 
So far only 
a few
sources (exclusively protoplanetary nebulae) 
are reported to have $I_{3.4}/I_{3.3}$\,$\gtsim$\,1 
(Hrivnak et al.\ 2007). They are atypical UIE sources: 
their UIE spectra have most of the power emitted from two broad bands 
peaking at $\simali$8$\mum$ and $\simali$11.8$\mum$, 
while typical UIE spectra have distinctive peaks at 7.7, 8.6, 
and 11.3$\mum$ (see Tokunaga 1997).\footnote{%
  Overall, the IR spectra of coals or kerogens 
  resemble that of atypical sources (e.g., some 
  protoplanetary nebulae; see Guillois et al.\ 1996).
  They do not resemble the UIE features
  seen in the interstellar sources except for highly 
  evolved (i.e., highly aromatized) coals
  (see Papoular 2001).
  }

We note that $N_{\rm C,aliph}/N_{\rm C,arom}$\,=\,0.09 
is an upper bound as the 3.4$\mum$ emission feature could 
also be due to anharmonicity of the aromatic C--H stretching 
mode (Barker et al.\ 1987).
Let $\nu$ be the vibrational quantum number. 
In a harmonic oscillator, the level spacing is constant; 
the $\Delta\nu$\,=\,1 transition between high $\nu$ levels 
results in the same spectral line 
as for the $\nu$\,=\,1$\rightarrow$0 transition. 
Anharmonicity decreases the spacing between the higher $\nu$ levels, 
and the $\Delta\nu$\,=\,1 transitions between higher $\nu$ levels 
occur at longer wavelengths. The anharmonicity model explains 
the weaker features (at 3.40$\mum$, 3.51$\mum$, ...) 
as ``hot bands'' 
($\nu$\,=\,2$\rightarrow$1, $\nu$\,=\,3$\rightarrow$2, ...)
of the 3.3$\mum$ fundamental $\nu$\,=\,1$\rightarrow$0
aromatic C--H stretching mode.
The 3.4$\mum$ emission feature could also be 
due
in part to ``superhydrogenated'' PAHs in which some peripheral 
C atoms have two H atoms (see Figure~2). 
The extra H atom converts the originally aromatic ring into 
an aliphatic ring. This creates two aliphatic C--H stretching bands: 
one due to the symmetric and the other 
to the asymmetric C--H stretching modes. 
These bands would fall near 3.4$\mum$ and 3.5$\mum$, 
with the former more intense than the latter, 
consistent with astronomical observations (Bernstein et al.\ 1996).
The 3.4$\mum$ feature may also result from aliphatic sidegroups 
attached as functional groups to PAHs 
(see Figure~2; Duley \& Williams 1981, Pauzat et al.\ 1999,
Wagner et al.\ 2000).
The C--H stretching modes of methyl (-CH$_3$), 
methylene (-CH$_2$-), and ethyl (-CH$_2$CH$_3$) 
sidegroups on PAHs fall near the weaker satellite features 
associated with the 3.3$\mum$ band. 
All these possibilities (i.e., anharmonicity, 
superhydrogenation, and aliphatic sidegroups) 
probably contribute to the 3.4$\mum$ emission, 
the extent of each depending on conditions 
in the local environment.
Sandford (1991) argued that the satellite features 
at 3.40, 3.46, 3.51, and 3.56$\mum$ in NGC\,7027 
cannot be predominantly due to aliphatic sidegroups on PAHs.

KZ11
note that the 3.4$\mum$ aliphatic 
C--H stretching mode is commonly observed in {\it absorption} 
in the diffuse ISM. If the UIE carriers have the same 
mixed aromatic-aliphatic structure as the bulk 
of the hydrocarbon material, then in heavily obscured regions, 
both the 3.3$\mum$ band and the 3.4$\mum$ band would show up 
in {\it absorption}, with the 3.4$\mum$ absorption band 
much {\it weaker} than the 3.3$\mum$ absorption band. 
However, astronomical observations have actually shown 
the opposite (see Figure~1): the 3.4$\mum$ absorption band 
is much {\it stronger} than the 3.3$\mum$ absorption band 
(e.g. in the Galactic center source GCS 3, 
the 3.4$\mum$ absorption band is stronger than 
the 3.3$\mum$ band by a factor of 35; Chiar et al.\ 2000). 
Therefore, the bulk of the 3.4$\mum$ absorber in the ISM 
must be hydrocarbon material in the larger grains, 
evidently more strongly aliphatic than the UIE carriers 
(Dartois et al.\ 2007).

\section{Constraints 
         from the 6.85$\mum$ Feature\label{sec:ch6.85um}}
In addition to the 3.4$\mum$ C--H stretching mode, 
aliphatic hydrocarbon materials also have two C--H 
deformation bands at 6.85$\mum$ and 7.25$\mum$.\footnote{%
   One may argue that in the KZ11-type coal- or kerogen-like
   material, the aliphatic C--H bands may not occur
   at the same wavelengths as for pure aliphatics 
   or PAHs with simple aliphatic sidegroups:  
   the aliphatic H atoms occupy a 
   broad range of 
   local chemical environments, subject to hydrogen 
   bonding perturbations by nearby O and S atoms. 
   Such interactions could conceivably shift the C--H frequencies 
   from their ``normal'' aliphatic positions. 
   However, laboratory measurements have shown that the aliphatic 
   C--H bands in coal or kerogen do occur at 3.4$\mum$ and 6.85$\mum$, 
   displaying little wavelength shift compared to that of pure aliphatics
   (see Papoular 2001).
   }
These two bands have been observed in weak absorption 
in the diffuse ISM (Chiar et al.\ 2000).
They are also seen in emission in interstellar 
and circumstellar UIE sources. 
Their strengths (relative to the nearby 7.7$\mum$ 
C--C stretching band) also allow an estimate of 
the aliphatic fraction of the UIE carrier.\footnote{%
   Coals or kerogens do not exhibit a distinct band 
   at 7.7$\mum$ and thus one cannot infer
   their aliphatic fractions from 
   $I_{6.85}/I_{7.7}$.
   }

\begin{figure}
\begin{center}
\includegraphics[width=16cm,angle=0]{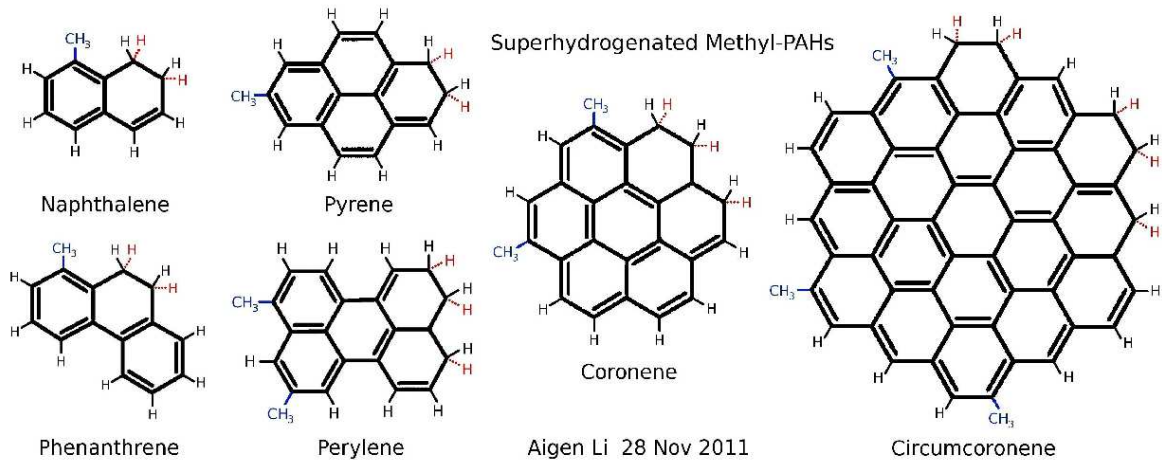}
\caption{\footnotesize
         \label{fig:fig2}
         Examples
         of ``superhydrogenated'' PAHs 
         with methyl (-CH$_3$) aliphatic sidegroups. 
         In addition to anharmonicity, superhydrogenation 
         and methyl-like aliphatic sidegroups attached to PAHs 
          may contribute to
         the weak 3.4$\mum$ 
         emission feature accompanying the 3.3$\mum$ feature.
         }
\end{center}
\end{figure}

Let $I_{6.85}$ and $I_{7.7}$ 
%
be the observed intensities 
of the 6.85$\mum$ and 7.7$\mum$ emission features. 
Let $A_{6.85}$ and $A_{7.7}$ 
%
be the strengths 
of the 6.85$\mum$ aliphatic C--H band 
and the 7.7$\mum$ aromatic C--C band. 
We take $A_{6.85} = 2.3\times10^{-18}\cm$ 
per CH$_2$ or CH$_3$ group,
an average of
that measured 
for methylcylcohexane 
($A_{6.85} = 3.0\times10^{-18}\cm$ per CH$_3$ group;
d'Hendecourt \& Allamandola 1986)
and for 
hydrogenated amorphous carbon 
($A_{6.85} = 1.5\times10^{-18}\cm$ 
per CH$_2$ or CH$_3$ functional group; 
Dartois \& Mu\~noz-Caro 2007).\footnote{%
  Typical type II Kerogens have
  $A_{6.85} \approx 3.6\times10^{-19}\cm$ per C atom 
  (Papoular 2001).
  If 15\% of the C atoms in kerogens are in aliphatic form,
  for kerogens we would have $A_{6.85} \approx 2.4\times10^{-18}\cm$ 
  per aliphatic C atom.  
  This is close to that adopted in this work:
  $A_{6.85} \approx 2.3\times10^{-18}\cm$ 
  per aliphatic C atom.  
  }

We take $A_{7.7} = 5.4\times10^{-18}\cm$ per C atom 
for charged aromatic molecules (Draine \& Li 2007).
Let $N^\prime_{\rm C,aliph}$ and $N^\prime_{\rm C,arom}$
respectively be the numbers of aliphatic and aromatic C atoms 
in the emitters of the 6--8$\mum$ UIE bands. 
Let $B_\lambda\left(T\right)\propto
\lambda^{-3}/\left[\exp\left(hc/\lambda kT\right)-1\right]$ 
be the Planck function at wavelength $\lambda$ and temperature $T$
(where $h$ is 
Planck's
constant, $c$ is the speed of light,
and $k$ is 
Boltzmann's 
constant), 
with $B_{6.85}/B_{7.7}\approx0.9\pm0.2$ 
for $330<T<1000\K$. 
Then $N^\prime_{\rm C,aliph}/N^\prime_{\rm C,arom}\approx
\left(I_{6.85}/I_{7.7}\right)\times\left(A_{7.7}/A_{6.85}\right) 
\times\left(B_{7.7}/B_{6.85}\right)\approx0.10$ 
for NGC 7027 ($I_{6.85}/I_{7.7}\approx0.039$) 
and $N^\prime_{\rm C,aliph}/N^\prime_{\rm C,arom}\approx0.14$
for the Orion bar ($I_{6.85}/I_{7.7}\approx0.053$). 
%
KZ11
derived $I_{6.85}/I_{7.7}\approx1.43$
for IRAS\,22272+5435, but the observed spectrum and 
decomposition fit support a significantly smaller value.

We conclude that the carriers of the 6--8$\mum$ UIE bands 
are predominantly aromatic, with $<$\,15\% of the C atoms 
in aliphatic form. The aliphatic fraction, while still small, 
appears to be higher than estimated for the 3.3--3.4$\mum$ band carriers, 
consistent with increased aromatization of the smallest particles, 
which are heated to the highest temperatures.

\section{Discussion\label{sec:discussion}}
KZ11 attribute the broad plateau emission 
around 8 and 12$\mum$ to the aliphatic component
of the UIE carreirs. They hypothesize that the clustering
of aromatic rings may break up the simple methyl- or 
methylene-like sidegroups and hence the aliphatic 
components may take many other forms 
(e.g., -CH=CH$_2$, -CH=CH-, C=CH$_2$, C=C-H).
They speculate that the in-plane and out-of-plane bending modes    
of these sidegroups may combine to form the plateau.
We note that the PAH model naturally accounts for the so-called 
``plateau'' emission through the combined wings of
the C--C and C--H bands.
We also note that the clustering of aromatic rings and
aliphatic chains would be accompanied by forming new C--C bonds
and losing H atoms. Laboratory measurements of coals
have shown that 
lowering
the H content leads to 
aromatization (see Papoular 2001).

KZ11
claim that the PAH hypothesis 
postulates that the UIE emission is excited exclusively 
by far-UV photons, and that this is inconsistent with 
observation of UIE emission in reflection nebulae 
excited by cool stars (Sellgren et al.\ 1990). 
However, Li \& Draine (2002) 
explicitly
considered 
the excitation of PAHs by longer-wavelength photons, 
and showed that the light from even relatively cool stars 
can 
excite UIE emission, consistent with observations. 
The excitation of PAHs by visible or even near-IR photons 
with wavelengths up to $\simali$1--2$\mum$ has been further 
experimentally verified (Mattioda et al.\ 2005). 
%
KZ11
note the constancy of UIE band ratios 
in regions (e.g. the Carina nebula) where the radiation intensity 
changes by orders of magnitude. This is {\it precisely} what one 
expects if the emission comes from single-photon heating of 
nanoparticles [see Figure 13 of Li \& Draine (2001), 
Figure 4b of Li \& Draine (2002), 
Figure 1f of Draine \& Li (2007)].

KZ11
note that of the more than 160 molecules 
identified in circumstellar and interstellar environments, 
none is a PAH. This is true, but also not surprising because 
the mid-IR UIE bands -- the major observational information -- 
are representative of functional groups and do not fingerprint 
individual PAH molecules.\footnote{%
   The far-IR bands are more sensitive to 
   the skeletal characteristics of a molecule,
   and hence are more diagnostic of 
   the molecular identity and more powerful for 
   chemical identification of unknown species.
   In principle, far-IR spectroscopy could be
   used to test the KZ11 hypothesis:
   the KZ11-type material has an extremely ``floppy''
   structure compared to the more rigid PAHs,
   and therefore there would be many low frequency 
   skeletal bends and very low-frequency 
   pseudo-rotations about bond axes.
   However, there is little information on
   the far-IR spectroscopy of coal or kerogen.  
   Even for PAHs, this information is very limited
   (e.g., see Joblin et al.\ 2009, Zhang et al.\ 2010).
   }
KZ11
argue that the carrier 
of the UIR features cannot be a ``pure aromatic compound''. 
Proponents of the identification of the astronomical UIE features 
as coming from PAHs do not claim that the emitting material is 
``pure aromatic compound'', as strictly defined by a chemist. 
The astronomical material may well include 
a {\it minor} aliphatic component, 
as well as defects, substituents (e.g., N in place of C), 
partial dehydrogenation, and sometimes superhydrogenation
(Tielens 2008). 
Some of the nanoparticles may be multilayer aggregates of PAHs.

KZ11
state that PAH molecules have strong 
and narrow absorption features in the UV whereas the search 
for characteristic absorption features of PAHs superposed 
on the interstellar extinction curves was not successful
(e.g., see Clayton et al.\ 2003). 
For individual {\it small} PAHs, this is true. 
However, in the PAH hypothesis it is natural to 
expect that there will be a large number of distinct 
species present in the ISM, and no single UV band may 
be strong enough to be identified in the UV. 
This also explains why laboratory-measured spectra 
of {\it individual} PAHs do not precisely match 
the observed UIE features in band widths and peak wavelengths, 
while {\it combined} laboratory spectra of neutral PAHs 
and their ions can successfully reproduce the UIE bands 
associated with many different interstellar objects 
(Allamandola et al.\ 1999).
There are, in fact, over 400 diffuse interstellar bands (DIBs) 
in the optical that remain to be identified 
(Sarre 2006, Salama et al.\ 2011).
Many of these may eventually be found to be produced 
by specific PAHs, but at this time we lack the laboratory 
spectroscopy to make the identifications. 
The lack of identification of any specific PAH 
is not a fatal problem for the PAH hypothesis, 
at least at this time. As we develop a better knowledge 
of the gas-phase spectroscopy of the larger PAHs, 
this story may change. 
If the DIBs are electronic transitions of PAHs, 
they hold great promise for identifying 
specific PAH molecules, as the electronic transitions 
are more characteristic of a specific PAH molecule 
than the mid-IR C--H and C--C vibrational bands.

The strong interstellar 217.5\,nm extinction bump 
is likely to be a {\it blend} of $\pi$\,--\,$\pi^{\ast}$ 
absorption bands from the entire population of PAHs, 
with the fine structures from individual PAH molecules smoothed out. 
Furthermore, internal conversion may lead to extreme broadening of 
the UV absorption bands in larger PAHs, which may account for 
absence of recognizable absorption features shortward of 200\,nm. 
This has been demonstrated both experimentally and theoretically. 
L\'eger et al.\ (1989) measured the absorption spectra of mixtures 
of over 300 neutral PAH species with $\simali$12--28 C atoms. 
Joblin et al.\ (1992) measured the absorption spectra of 
neutral PAH mixtures containing $\simali$14--44 C atoms. 
All these spectra have a strong UV feature around 217.5\,nm 
(but relatively broader than the interstellar bump). 
While it is true that 
laboratory PAH samples
do not precisely reproduce the observed profile of 
the 217.5\,nm feature (L\'eger et al.\ 1989, Joblin et al.\ 1992), 
this is probably due to the fact that laboratory studies 
are generally limited to small PAHs while interstellar PAHs 
are much larger (e.g. models that reproduce 
the $\simali$3--20$\mum$ UIE bands have most of 
the PAH mass in PAHs with $>$\,100 C atoms, 
see Li \& Draine 2001, Draine \& Li 2007). 
Indeed, Steglich et al.\ (2010) showed that larger PAHs 
indeed provide better fits to the observed 217.5\,nm feature. 
Cecchi-Pestellini et al.\ (2008) also showed that a weighted 
sum of 50 neutral and ionized PAHs in the size range of 
$\simali$10--66 C atoms 
can
reproduce the 217.5\,nm 
extinction bump observed in various environments.

\section{Conclusion}\label{sec:summary}
We examine the hypothesis of mixed aromatic-aliphatic 
organic matter as the UIE carriers.
We place an upper limit on the aliphatic fraction of 
the UIE carriers based on the observed
weak
intensities of 
the 3.4$\mum$ and 6.85$\mum$ emission features.
By attributing them {\it exclusively} to 
aliphatic C--H stretch and aliphatic C--H deformation,
we derive the fraction of carbon atoms
in aliphatic form to be $<$\,15\%.
We conclude that the UIE emitters are predominantly 
aromatic: PAHs
%
dominate the principal UIE bands.
Our expectation is that confirmation will not come 
until we have laboratory spectroscopy of PAH candidates 
in the gas phase that precisely match some of the observed DIBs. 

\acknowledgments{%
AL is supported in part by NSF AST-1109039. 
BTD is supported in part by NSF AST-1008570.
We thank R.\ Glaser, M.\ K\"ohler, S.\ Kwok,
and the anonymous referee for helpful comments.
}



\begin{thebibliography}{30}
\expandafter\ifx\csname natexlab\endcsname\relax\def\natexlab#1{#1}\fi
\bibitem[]{}Allamandola, L.J., Tielens, A.G.G.M., \& Barker, J.R.\ 
            1985, ApJ, 290, L25
\bibitem[]{}Allamandola, L.J., Hudgins, D.M., \& Sandford, S.A.\ 
            1999, ApJ, 511, 115
\bibitem[]{}Barker, J.R., Allamandola, L.J., \& Tielens, A.G.G.M.\
            1987, ApJ, 315, L61
\bibitem[]{}Bernstein, M.P., Sandford, S.A., \& Allamandola, L.J.\
            1996, ApJ, 472, L127
\bibitem[]{}Cecchi-Pestellini, C., Malloci, G., Mulas, G., 
            Joblin, C., \& Williams, D.A.\ 
            2008, A\&A, 486, L25 
\bibitem[]{}Chiar, J.E., Tielens, A.G.G.M., Whittet, D.C.B., et al.\
            2000, ApJ, 537, 749
\bibitem[]{}Clayton, G.C., et al.\ 2003, ApJ, 592, 947
\bibitem[]{}Cody, G.D., et al.\ 2011,
            PNAS, 108, 19171
\bibitem[]{}Dartois, E., \& Mu\~noz-Caro, G.M.\
            2007, A\&A, 476, 1235
\bibitem[]{}Dartois, E., Geballe, T.R., Pino, T., et al.\
            2007, A\&A, 463, 635
\bibitem[]{}d'Hendecourt, L.B., \& Allamandola, L.J.\
            1986, A\&AS, 64, 453
\bibitem[]{}Derenne, S., \& Robert, F.\ 2010,
            Meteorit. Planet. Sci., 45, 1461
\bibitem[]{}Draine, B.~T., \& Li, A. 2007, ApJ, 657, 810
\bibitem[]{}Duley, W.W., \& Williams, D.A.\ 1981, MNRAS, 196, 269
\bibitem[]{}Duley, W.W., \& Williams, D.A.\ 2011, ApJ, 737, L44
\bibitem[]{}Geballe, T.R., Lacy, J.H., Persson, S.E., 
            McGregor, P.J., \& Soifer, B.T.\
            1985, ApJ, 292, 500
\bibitem[]{}Goto, M., Usuda, T., Takato, N., et al.\
            2003, ApJ, 589, 419
\bibitem[]{}Guillois, O., Nenner, I., Papoular, R., 
            \& Reynaud, C.\ 1996, ApJ, 464, 810
\bibitem[]{}Hrivnak, B.J., Geballe, T.R., \& Kwok, S.\
            2007, ApJ, 662, 1059
\bibitem[]{}Joblin, C., Bern\'e, O., Simon, A., \& Mulas, G.\ 
            2009, Cosmic Dust -- Near and Far (ASP Conf. Ser. 414), 
            ed. Th. Henning, E. Gr\"un, \& J. Steinacker 
            (San Francisco, CA: ASP), 383 
\bibitem[]{}Joblin, C., \& Tielens, A.G.G.M.\
            2011, PAHs and the Universe: A symposium to 
            Celebrate the 25th Anniversary of the PAH Hypothesis, 
            EAS Publ. Ser., 46
\bibitem[]{}Joblin, C., L\'{e}ger, A., \& Martin, P. 1992, 
            ApJ, 393, L79
\bibitem[]{}Joblin, C., Tielens, A.G.G.M., Allamandola, L.J., 
            \& Geballe, T.R.\ 1996, ApJ, 458, 610
\bibitem[]{}Jourdain de Muizon, M., Geballe, T.R., d'Hendecourt, L.B., 
            \& Baas, F.\ 1986, ApJ, 306, L105
\bibitem[]{}Kondo, T., Kaneda, H., Oyabu, S., et al.\ 
            2012, ApJ, 751, L18 
\bibitem[]{}Kwok, S., \& Zhang, Y.\
            2011, Nature, 479, 80 (KZ11)
\bibitem[]{}L\'{e}ger, A., \& Puget, J. 
            1984, A\&A, 137, L5
\bibitem[]{}L\'{e}ger, A., Verstraete, L., d'Hendecourt, L., 
            D{\'e}fourneau, D., Dutuit, O., Schmidt, W., 
            \& Lauer, J.\ 1989, in Interstellar Dust (IAU Symp.\,135),
 	    ed. L.J. Allamandola \& A.G.G.M. Tielens 
            (Dordrecht: Kluwer), 173
\bibitem[]{}Li, A., \& Draine, B.T.\ 2001, 
            ApJ, 554, 778
\bibitem[]{}Li, A., \& Draine, B.T.\ 2002, 
            ApJ, 572, 232
\bibitem[]{}Mattioda, A.L., Allamandola, L.J., \& Hudgins, D.M.\
            2005, ApJ, 629, 1183
\bibitem[]{}Mu\~noz-Caro, G.M., Ruiterkamp, R., Schutte, W.A., 
            Greenberg, J.M., \& Mennella, V.\ 2001,
            A\&A, 367, 347
\bibitem[]{}Papoular, R.\ 2001, A\&A, 378, 597
\bibitem[]{}Pauzat, F., Talbi, D., \& Ellinger, Y.\
            1999, MNRAS, 304, 241
\bibitem[]{}Pendleton, Y.J., \& Allamandola, L.J.\
            2002, ApJS, 138, 75
\bibitem[]{}Puget, J.L., L\'eger, A., \& d'Hendecourt, L.\
            1995, A\&A, 293, 559
\bibitem[]{}Salama, F., Galazutdinov, G.~A., Kre{\l}owski, J., 
            Biennier, L., Beletsky, Y., \& Song, I.-O.\ 2011, 
            ApJ, 728, 154 
\bibitem[]{}Sandford, S.A.\ 1991, ApJ, 376, 599
\bibitem[]{}Sarre, P.J.\ 2006, J. Mol. Spec., 238, 1 
\bibitem[]{}Schutte, W.A., Tielens, A.G.G.M., 
            \& Allamandola, L.J.\ 1993, ApJ, 415, 397
\bibitem[]{}Sellgren, K., Luan, L., \& Werner, M.W.\
            1990, ApJ, 359, 384
\bibitem[]{}Sloan, G.C., Bregman, J.D., Geballe, T.R., 
            Allamandola, L.J., \& Woodward, C.E.\
            1997, ApJ, 474, 735
\bibitem[]{}Steglich, M., J{\"a}ger, C., Rouill{\'e}, G., 
            Huisken, F., Mutschke, H., \& Henning, T.\ 
            2010, ApJ, 712, L16 
\bibitem[]{}Tielens, A. G. G. M. \ 2008, ARA\&A, 46, 289
\bibitem[]{}Tokunaga, A.T.\ 1997, in ASP Conf. Ser. 124, 
            Diffuse Infrared Radiation and the IRTS, 
            ed. H. Okuda, T. Matsumoto, \& T. Roellig 
            (San Francisco, CA: ASP), 149 
\bibitem[]{}Wagner, D.R., Kim, H., \& Saykally, R.J.\
            2000, ApJ, 545, 854
\bibitem[]{}Zhang, J., Han, F., Pei, L., Kong, W., 
            \& Li, A.\ 2010, ApJ, 715, 485 
\bibitem[]{}Yamagishi, M., Kaneda, H., Ishihara, D., et al.\ 
            2012, A\&A, 541, A10 
%
\end{thebibliography}
\end{document}